\documentclass[aps,prl,twocolumn,showpacs,amsmath,floatfix,superscriptaddress]{revtex4}

\usepackage{graphicx}

\begin{document}

\title{Quasiparticle Energies and Band Gaps of Graphene Nanoribbons}

\author{Li Yang}
\affiliation{Department of Physics, University of California at
Berkeley, California 94720} \affiliation{Materials Sciences
Division, Lawrence Berkeley National Laboratory, Berkeley,
California 94720}

\author{Cheol-Hwan Park}
\affiliation{Department of Physics, University of California at
Berkeley, California 94720} \affiliation{Materials Sciences
Division, Lawrence Berkeley National Laboratory, Berkeley,
California 94720}

\author{Young-Woo Son}
\affiliation{Department of Physics, Konkuk University, Seoul
143-701, Korea}

\author{Marvin L. Cohen}
\affiliation{Department of Physics, University of California at
Berkeley, California 94720} \affiliation{Materials Sciences
Division, Lawrence Berkeley National Laboratory, Berkeley,
California 94720}

\author{Steven G. Louie}
\affiliation{Department of Physics, University of California at
Berkeley, California 94720}
\affiliation{Materials Sciences Division, Lawrence Berkeley
National Laboratory, Berkeley, California 94720}

\date{\today}

\begin{abstract}

We present calculations of the quasiparticle energies and band
gaps of graphene nanoribbons (GNRs) carried out using a
first-principles many-electron Green's function approach within
the GW approximation. Because of the quasi-one-dimensional nature
of a GNR, electron-electron interaction effects due to the
enhanced screened Coulomb interaction and confinement geometry
greatly influence the quasiparticle band gap. Compared with
previous tight-binding and density functional theory studies, our
calculated quasiparticle band gaps show significant self-energy
corrections for both armchair and zigzag GNRs, in the range of
0.5-3.0 eV for ribbons of width 2.4-0.4 nm. The quasiparticle band
gaps found here suggest that use of GNRs for electronic device
components in ambient conditions may be viable.
\end{abstract}


\maketitle


Graphene, a single atomic layer of graphite, has been successfully
produced in experiment \cite{novoselov04,novoselov05,kim05}, which
has resulted in intensive investigations on graphene-based
structures because of fundamental physics interests and promising
applications
\cite{novoselov05p,kim06,zhang05,berger05,neto06,mele05,neto05}.
When graphene is etched or patterned along one specific direction,
a novel quasi-one-dimensional structure, a strip of graphene of
nanometers in width, can be obtained which is referred to as a
graphene nanoribbon (GNR). The GNRs are predicted to exhibit
various remarkable properties and may be a potential elementary
structure for future carbon-based nanoelectronics
\cite{yw06,waka01,scuseria06,white07}. In particular, as a
fundamental factor in determining transport and optical
properties, the electronic band structure of GNRs has been the
subject of great interest.

Depending on specific GNRs, previous studies using tight-binding
or massless Dirac fermion equation approaches have predicted GNRs
to be either metals or semiconductors
\cite{nakada96,waka99,ezawa06,brey06,saito06,levitov06}. Whereas,
density functional theory (DFT) calculation showed that all
zigzag-edged and armchair-edged GNRs have a finite band gap when
relaxation of the structure or spin polarization is considered
\cite{ywprl06,scuseria06}. Recent experiments have reported finite
band gaps in all the GNRs that have been tested
\cite{avouris07,kim07}. However, it is well established
\cite{louiebook} that the Kohn-Sham eigenvalues from DFT
calculation are inappropriate to describe the band gaps of
semiconductors. The disagreement between the Kohn-Sham band gap
and experimental data is worse for nanostructures because of the
enhanced electron-electron interaction in those systems. On the
other hand, first-principles calculation based on many-body
perturbation theory, such as the GW approximation
\cite{louiebook,steve86}, has been shown to be reliable for
obtaining quasiparticle band gaps of nano-sized semiconductors
\cite{catalin04, zhao04, rubio06, CH06}. Motivated by the
importance but the lack of accurate knowledge about quasiparticle
band gaps of the GNRs and by the successes of the GW approximation
for nano-size semiconductors, we carry out a first-principles
calculation using the GW approximation to determine the
quasiparticle energy spectrum and the band gaps of the GNRs.

\begin{figure}
\includegraphics*[scale=0.45]{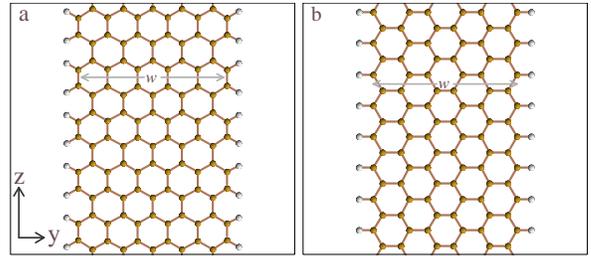}
\caption{\label{fig:struc} (Color online) (a) A ball-stick model
for an 11-AGNR which has 11 C-C dimer lines making up its width
$w$. Hydrogen atoms (white balls) are used to passivate the edge
$\sigma$-dangling bonds. $x$, $y$ and $z$ are the Cartesian
coordinates. (b) A ball-stick model for a 6-ZGNR which has 6
zigzag chains along the $z$ direction.}
\end{figure}

We consider two common types of GNRs. Their structures are shown
in Fig.~\ref{fig:struc}. The left one, called armchair GNR (AGNR),
has armchair-shaped edges; the right one, called zigzag GNR
(ZGNR), has zigzag-shaped edges. The dangling $\sigma$-bonds at
the edges are passivated by hydrogen atoms. The structures of the
GNRs studied here are fully relaxed according to the forces and
stress on the atoms. Following conventional notation, a GNR is
specified by the number of dimer lines or zigzag chains along the
ribbon forming the width, for the AGNR and ZGNR respectively, as
explained in Fig.~\ref{fig:struc}. For example, the structure of
Fig.~\ref{fig:struc} (a) is referred as a 11-AGNR and the
structure in Fig.~\ref{fig:struc} (b) as a 6-ZGNR. In addition,
when referring to the width of a GNR here, we define the width
without including the hydrogen atoms at the edge, as shown in
Fig.~\ref{fig:struc}.

Following the approach of Hybertsen and Louie \cite{steve86}, we
first obtain the electronic ground state with DFT within the local
(spin) density approximation [L(S)DA]. Then, the quasiparticle
energies are calculated within the GW approximation to the
electron self energy. Norm-conserving pseudopotentials
\cite{troullier91} and the plane-wave basis are used. In this
calculation, k-grid is sampled uniformly along the 1-D Brillouin
zone. To assure that the quasiparticle energies are converged to
within 0.1 eV, a 1x1x32 k-point sampling is used for AGNRs and a
1x1x64 k-point sampling for ZGNRs. Since the supercell method is
used in this calculation to mimic isolated GNRs, we use a
truncated Coulomb interaction to eliminate the image effect
between adjacent supercells \cite{catalinapl,trunc1,trunc2}.
Considering the geometry of the ribbons, we employ a
rectangular-shape Coulomb truncation as
\begin{equation}\label{1}
V_{c}=\frac{1}{r}\,\theta(|x|-x_c)\,\theta(|y|-y_{c})\,\theta(|z|-z_{c}),
\end{equation}
where $r=\sqrt{x^2+y^2+z^2}$ is the distance between two
electrons; $x_c$, $y_c$ and $z_c$ are cutoff parameters. As
discussed in previous studies \cite{catalinapl}, the dimension of
the unit cell has to be $2x_c \times 2y_c \times 2z_c$. Because of
the single layer structure of GNRs, the truncation lengths $x_c$
and $z_c$ are fixed for all GNRs in our calculations. The unit
cell volume is linearly dependent on the width of a GNR, and the
number of plane waves needed is also scaled linearly with the
width of ribbon, which significantly reduces the cost of the
computation.

Another important aspect of the calculation is that we have to
include the spin degree of freedom to account for the spin
polarization in ZGNRs. It is shown that the static polarizability
matrix is diagonal in spin space \cite{li02}. Combining this with
the fact that the bare Coulomb interaction is independent of the
spin degree of freedom, the spin-polarized GW calculation proves
to be almost the same as the non-spin-polarized case with the
exception of replacing the static polarizability with the sum of
its two diagonal spin components. The details of the
spin-polarized GW calculation can be found in Ref\cite{li02}.

\begin{figure}
\includegraphics*[scale=0.92]{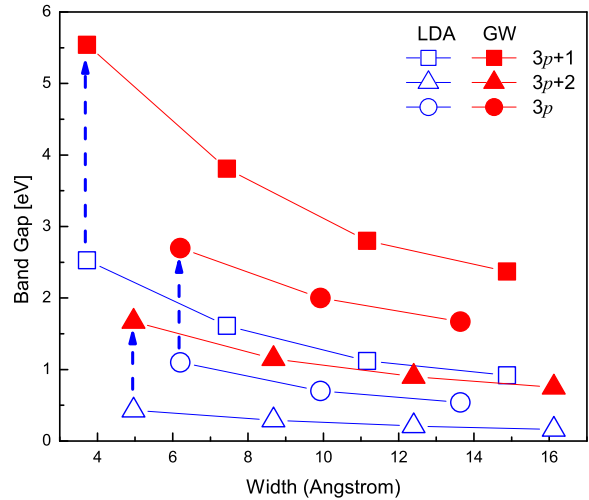}
\caption{\label{fig:gw-a} (Color online) Variation of band gaps
with the width of AGNRs. The three families of AGNRs are
represented by different symbols. The values of the same family of
AGNRs are connected by solid lines as guides to the eyes. The open
symbols are LDA band gaps while the solid symbols are the
corresponding quasiparticle band gaps. Dashed arrows are used to
indicate the self-energy correction for the smallest width ribbon
of each of the three families of AGNRs studied.}
\end{figure}

The LDA and quasiparticle band gaps of eleven armchair GNRs are
shown in Fig.~\ref{fig:gw-a}. As is found in the LDA, the
quasiparticle band structure has a direct band gap at the zone
center for all AGNRs studied. In addition, the band gaps of the
three families of n-AGNRs, which are classified according to
whether n=$3p+1$,$3p+2$ or $3p$ (n is the number of dimer chains
as explained in Fig 1, and $p$ is an integer), show qualitatively
the same hierarchy as those obtained in LDA
$(E_g^{3p+1}>E_g^{3p}>E_g^{3p+2}\neq 0)$.

However, the GW self-energy corrections to the band gap, $E_g$,
are significant for all the AGNRs. The corrections are from 0.5 to
3 eV for the AGNRs in Fig.~\ref{fig:gw-a} with width from 1.6 to
0.4 nm, which are much larger than those found for bulk graphite
or diamond \cite{steve86}. A weaker screening contributes to this
enhanced self-energy correction because the GNRs are isolated and
surrounded by vacuum that does not screen the Coulomb interaction.
In addition, the confined geometry (one-dimensional nature) of the
GNRs enhances the effect of electron-electron interaction, which
further enlarges the self-energy correction. This kind of enhanced
self-energy correction is also found in other nanostructures such
as nanotubes and nanowires \cite{catalin04, zhao04, rubio06,
CH06}.

The band gaps from both the LDA and GW calculations show clearly
size dependence in Fig.~\ref{fig:gw-a} because of quantum
confinement. Under a hard-wall boundary condition used in previous
works, an inverse relation, $E_g \propto 1/(w+w_0)$, is widely
applied to characterize the size dependence of the band gap in
AGNRs, where $w$ is the width defined in Fig.~\ref{fig:gw-a} and
$w_0$ is a small constant (2.4 \AA). This size-dependence of the
band gaps describes the tight-binding and LDA results well [21].
However, the boundary condition for the GNRs is not strictly a
hard wall condition, and the electron distribution will leak out
of the boundary more or less. Therefore the effective width of
GNRs should be larger than the physical width $w$. Considering
this effect, we use the formula
\begin{equation}\label{2}
E_g = \frac{a}{w+w_0+\delta},
\end{equation}
to fit the band gap values in Fig.~\ref{fig:gw-a}, and the fitted
results are given in Table I. For LDA results, the parameter
$\delta$ is close to zero or a little bit negative. For GW
results, the parameter $\delta$ found is around 1.5 to 2.9 $\AA$.
Therefore the correction to the effective width for each edge is
only around 0.75 to 1.45 $\AA$, which may reflect the
non-hard-wall nature of the confinement. In the limit of wider
GNRs, $\delta$ and $w_0$ can be ignored, the trend reduces to $E_g
\propto 1/w$.

\begin{table}
\caption{\label{table:fit} Fitted parameters for the LDA and GW
band gaps of AGNRs according to formula (2). The value of fitted
$\delta$ reflects an effective width correction.}
\begin{ruledtabular}
\begin{tabular}{ccccc}
   &    LDA   &  &  GW  &  \\
\hline
 family & a(eV $\cdot$ \AA)  & $\delta$(\AA)  & a(eV $\cdot$ \AA)  & $\delta$(\AA)\\
\hline
 &  &  &  & \\
3\emph{p}+1   & 15.8  &    0.1 &  44.4  & 1.8  \\
3\emph{p}+2  & 3.0  & -0.4 & 14.6 & 1.3 \\
3\emph{p} &     7.6   & -1.7 & 31.3 & 2.9 \\
\end{tabular}
\end{ruledtabular}
\end{table}

\begin{figure}
\includegraphics*[scale=0.75]{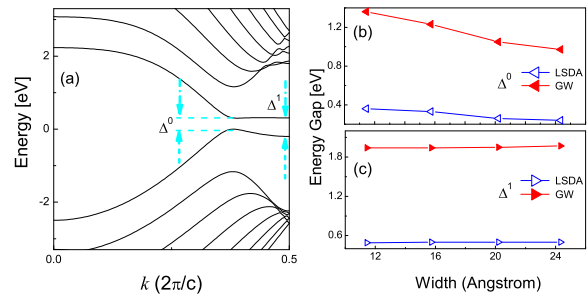}
\caption{\label{fig:gw-z} (Color online) Calculated band structure
and energy gap of ZGNRs. (a) The LSDA band structure of a 12-ZGNR.
The up and down spin states are degenerated for all the bands, and
the top of the valence band is set at zero. The symbols,
$\Delta^0$ and $\Delta^1$ denote the direct band gap and the
energy gap at the zone boundary. (b) Variation of direct band gap
with the width of ZGNRs. The open symbols denote the LSDA results
while the solid symbols are the GW results. (c) Variation of the
energy gap at the zone boundary with the width of ZGNRs. The
symbols have the same meaning as those in (b).}
\end{figure}

Fig.~\ref{fig:gw-z} (a) shows the LSDA bandstructure of 12-ZGNR.
There are two notable characteristics in the electronic structure
of ZGNRs: 1) the top of valence band and the bottom of conduction
band are composed of mainly edge states; and 2) the spin
interaction introduces a finite band gap in the ZGNRs. As shown in
Fig.~\ref{fig:gw-z} (b) and (d), the self-energy corrections to
the LSDA energy gaps in ZGNRs are similar with those in AGNRs, and
the corrections enlarge the band gap by 0.8 eV to 1.5 eV for the
ribbons studied. The spin polarization changes the screening type
of ZGNRs from that of a metal to that of a semiconductor.
Therefore a significant self-energy correction is resulted as in
the case of the AGNRs. Based on the discussion of AGNRs, we also
try to fit the width dependence of the quasiparticle band gaps in
Fig.~\ref{fig:gw-z} (b). We fit the results directly with a
functional form of $1/(w+\delta)$. The fitted $\delta$ of LSDA is
almost zero, and it is 16 $\AA$ for the GW values, which is much
larger than that in AGNRs. This is not unexpected, because it is
the spin interaction between electrons close to the edge that
induces the finite band gap in ZGNRs. Therefore we do not expect a
simple quantum-confinement effect, a $1/w$ size-dependence of the
band gap, in such narrow ZGNRs.

Unlike the band gap ($\Delta^0$) located around three-fourth of
the way to the Brillouin zone edge (Fig.~\ref{fig:gw-z} (a)), the
energy gap at the zone boundary ($\Delta^1$) is not sensitive to
the width of ZGNRs as seen in Fig.~\ref{fig:gw-z} (c). Previous
tight-binding calculations [33] show that the profile of edge
states decays to the center of ZGNRs with the factor of $e^{-ar}$,
where $a=-\frac{2}{\sqrt{3}c}$ln$|2$cos$\frac{kc}{2}|$
($\frac{2\pi}{3} \leq kc \leq \pi, c$ is the lattice constant of
ZGNRs along the $z$ direction). As a result, the band edge states
close to the zone boundary are highly confined at the edge of
ZGNRs. Because of their dominant edge-state character, these
states are not sensitive to the width of the ribbons, hence the
gap $\Delta^1$ is virtually independent of width.

\begin{figure}
\includegraphics*[scale=0.35]{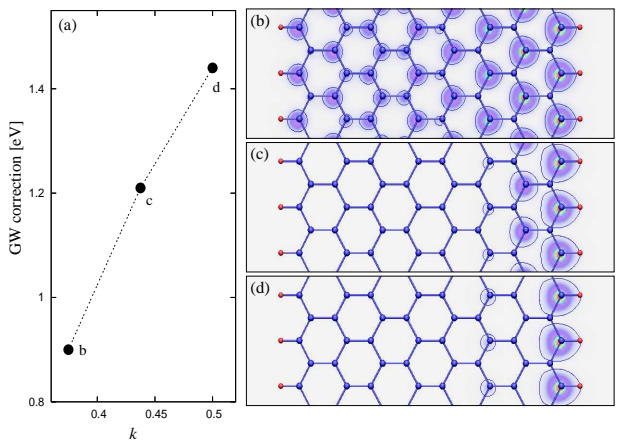}
\caption{\label{fig:gw-wf} (Color online) (a) Variation of GW
correction (the difference between the quasiparticle gap and the
LSDA gap) with wavevector of electronic states ($k$ = 0.375,
0.4375 and 0.50 in units of $2\pi/c$) in an 8-ZGNR. (b), (c) and
(d) are the charge distributions of the conduction state with the
corresponding wavevectors in (a). Because the up and down spin
components are degenerated but localized on a different
sublattice, we plot the charge distribution of only one spin
component.}
\end{figure}

Since the electronic wavefunction of the edge states is more and
more confined to an edge of a ZGNR when its wavevector $k$
approaches the zone boundary, it provides a possibility to see how
the self-energy correction evolves with the localization of the
electronic state. We plot the charge distributions of three
electronic states of the first conduction band with different
wavevector $k$ and their corresponding self-energy correction
values defined as $E^{QP}-E^{LSDA}$ in Fig.~\ref{fig:gw-wf}. It is
clear that the self-energy correction is enhanced when the state
is confined at the edge as shown from Fig.~\ref{fig:gw-wf} (b) to
(d). Because of the $1/r$ nature of the Coulomb interaction, the
self-energy of a state is sensitive to the localization of the
wavefunction. Therefore a larger self-energy correction is found
to the more localized edge state. As a consequence, the dependence
of the GW correction on wavevector significantly changes the band
dispersion in ZGNRs from that of LSDA calculations. A smaller
effective mass and better mobility for the carriers are expected
in ZGNRs for the GW bands as compared to the LSDA ones.

Recently, several experiments related to the quasiparticle band
gap in GNRs have been reported \cite{avouris07,kim07}. They have
not only proven the existence of finite band gap in GNRs but also
shown a larger gap when the width of the GNR decreases. Within a
range of width of GNRs of 15 nm to 90 nm, a $E_g \propto 1/w$
relation is observed. This finding agrees qualitatively with our
GW results. However, the experimental data are for the wider GNRs
where the widths are far from the range of widths of our
calculated GNRs (0.4-2.4 nm). In addition, all the GNRs in the
experimental case are etched by the oxygen plasma, which could be
different from our hydrogen passivated GNRs. Therefore it is
difficult to compare our GW results with current experimental data
directly. On the other hand, considering that the origin of the
enhancement of the self-energy correction in GNRs is the
quasi-one-dimensional geometry and weakened screening, we expect
that other passivating atoms or molecules do not change the
physics here significantly. With advance in experimental
techniques, it is very possible that smaller-sized and
hydrogen-passivated GNRs will soon be fabricated. A comparison
between our first-principles results and experimental data can
then be made.

In conclusion, we have performed a first-principles Green's
function calculation within the GW approximation to obtain the
quasiparticle band gaps in GNRs. Due to the enhanced
electron-electron interaction in these quasi-one-dimensional
systems, a significant self-energy correction is found for both
armchair and zigzag GNRs. The quasiparticle energy of states near
the band gap in ZGNRs is found to be wavevector sensitive, and
this gives rise to a larger band width and smaller effective mass
for carriers in ZGNRs. The calculated quasiparticle band gaps are
within the most interesting range (1-3 eV for 2-1 nm GNRs) and
give promise for applications of GNRs in nanoelectronics.

We thank F. Giustino, D. Prendergast and E. Kioupakis for
discussions. This research was supported by NSF Grant No.
DMR04-39768 and by the Director, Office of Science, Office of
Basic Energy under Contract No.DE-AC02-05CH11231. Y.-W. Son
acknowledges a support by the KOSEF grant funded by the MOST No.
R01-2007-000-10654-0. Computational resources have been provided
by Datastar at the San Diego Supercomputer Center.


\end{document}